\documentclass[preprint,prb]{revtex4}

\usepackage[english]{babel}
\usepackage[intlimits]{amsmath}
\usepackage[dvips]{graphicx}

\begin{document}

\begin{titlepage}
\
\title{Thermodynamics and Phase Transitions of Electrolytes on Lattices with Different Discretization Parameters}
\author{Maxim N. Artyomov}
\affiliation{Department of Chemistry, University of Chicago, Chicago, IL 60637, USA}    
\author{Anatoly B. Kolomeisky}
\affiliation{Department of Chemistry, Rice University, Houston, TX 77005, USA}

\begin{abstract}
Lattice models are crucial for studying  thermodynamic properties in many physical, biological and chemical systems. We investigate Lattice Restricted Primitive Model (LRPM) of electrolytes with different discretization parameters in order to understand thermodynamics  and the nature of  phase transitions in the systems with charged particles. A discretization parameter is defined as a number of lattice sites that can be occupied by each particle, and it allows to study the transition from the discrete picture to the continuum-space description. Explicit analytic and numerical calculations are performed using lattice Debye-H\"{u}ckel approach, which takes into account the formation of dipoles, the dipole-ion interactions  and correct lattice Coulomb potentials. The gas-liquid phase separation is found at low densities of charged particles for different  types of lattices. The increase in the  discretization parameter lowers the critical temperature and  the critical density, in agreement with Monte Carlo computer simulations results. In the limit of infinitely large discretization  our results approach the predictions from the continuum model of electrolytes. However, for the very fine discretization, where  each particle can only occupy  one lattice site,  the gas-liquid phase transitions are suppressed by order-disorder phase transformations.

\end{abstract}

\maketitle

\end{titlepage}

\section {Introduction}

Electrostatic interactions are important in various physical, chemical and biological processes.{\cite{levin02} However, a full thermodynamic description of systems with charged particles is still far from complete. In last decade this subject has attracted a lot of attention due to controversial theoretical and experimental issues on the nature of phase transitions in ionic fluids.\cite{fisher94,stell95,weingartner00}

Lattice models have been used extensively for investigations of  different phenomena in chemistry, physics and biology. For example, the  Ising model, which is a lattice gas model, is fundamental for understanding critical phenomena in non-charged systems. This observation has strongly stimulated many studies of ionic fluids utilizing the discretized lattice models.\cite{panagiotopoulos99,dickman99,brognara02,KKF02,kobelev02,panagiotopoulos02,ciach03,artyomov03,moghaddam03,kim04,ciach04,ciach04a}  There are several advantages of using lattice approach for systems with the particles interacting via  long-ranged Coulomb potentials. The production of ion pairs and  dipole-ion interactions can be described better than in  continuum-space models.\cite{KKF02,artyomov03} Lattice models are also computationally  much more efficient with respect to the time and length scales.\cite{panagiotopoulos99,panagiotopoulos02,moghaddam03,kim04} Monte Carlo computer simulations of charged systems on discretized lattices are faster by a factor of 5-100 than the corresponding continuum models.\cite{panagiotopoulos02,panagiotopoulos99,kim04}

If we define a diameter of a charged particle as $\sigma$ and the lattice cell size as $a$, then a discretization parameter $\zeta=\sigma / a$ specifies how close the lattice system approaches the continuum behavior. The case of $\zeta=1$ corresponds to the standard lattice model of electrolytes  where each particle occupies no more than one lattice site. When the discretization parameter  becomes very large the system of charged particles does not feel anymore the underlying lattice and its thermodynamic properties become indistinguishable from the properties of  continuum  electrolytes.   

Theoretical investigations of thermodynamics and phase behavior of lattice models of electrolytes have followed several different approaches.\cite{dickman99,brognara02,KKF02,kobelev02,ciach03,artyomov03,ciach04,ciach04a} The hierarchical reference theory\cite{brognara02} utilized the  renormalization group methods to calculate thermodynamic properties of ionic fluids on lattices. Ciach and coworkers\cite{ciach03,ciach04, ciach04a} used the field-theoretical methods to analyze different properties of lattice electrolytes. However these methods are mathematically very complicated and the predictions, as compared with Monte Carlo simulations,\cite{panagiotopoulos99,kobelev03} are mostly of qualitative nature. A different theoretical method\cite{KKF02,kobelev02,kobelev03,artyomov03} is based on the physically more transparent Debye-H\"{u}ckel (DH) approach. In this method the free energies of  lattice electrolytes are obtained by solving the lattice versions of usual Debye-H\"{u}ckel equations  for general dimensions. This method can provide a reasonable thermodynamic description of different lattice models of charged particles, such as charge-asymmetric\cite{artyomov03} and anisotropic lattices,\cite{kobelev02,kobelev03} as compared with computer simulations and other theoretical approaches.\cite{ciach04,ciach04a}

A full thermodynamic investigation of  continuum ($\zeta=\infty$) restricted primitive model (RPM), which is a system of equal-size hard-sphere  ions carrying positive and negative charges, using Debye-H\"{u}ckel  approach has shown that in the system there are  gas-liquid phase transitions driven by electrostatic interactions.\cite{FL96} However, a systematic study of the lattice version of RPM\cite{KKF02} (with $\zeta=1$) indicates that these phase transitions are suppressed on simple-cubic (sc)  and body-centered cubic (bcc) lattices because of thermodynamically more favorable charge-ordering phase transformations. These theoretical observations are in agreement with the predictions from Monte Carlo computers simulations.\cite{panagiotopoulos99,kobelev03}  The different thermodynamic properties  of lattice and continuum models of electrolytes raised the question of crossover phase behavior of ionic systems  with intermediate lattice discretization.  Ciach and Stell,\cite{ciach03,ciach04,ciach04a} using mean-field arguments and renormalization group methods, discussed some features of phase transitions for these lattice models. However, full thermodynamic and phase properties of these systems are not investigated yet. The goal of this article is to fill this gap.

In this paper we study the lattice models of charged particles with different discretization parameters by applying  Debye-H\"{u}ckel approach. In addition, we modify the original lattice model of electrolytes\cite{KKF02} ($\zeta=1$) that had several unphysical features. The paper is organized in the following way. In Sec. II we provide a general description of lattice models of electrolytes using the Debye-H\"{u}ckel method. Specific calculations and discussions of thermodynamic properties of charged systems for  lattices with different discretization parameters are given in Sec. III. Our results and conclusions are summarized in Sec. IV.

\section{General Lattice DH Theory}

Consider the Lattice Restricted Primitive Model (LRPM) of electrolytes  in three dimensions. Note that our method can be easily generalized in $d$ dimensions,\cite{KKF02} however, for clarity we prefer to explain the details using three-dimensional lattices.  The model consists of equal number of oppositely charged particles, with charges $q_{0}$ or $-q_{0}$, correspondingly. Due to the overall neutrality, the densities of positive and negative ions are equal to each other, $\rho_{+}=\rho_{-}=\frac{1}{2} \rho_{1}$. The particles have a spherical shape with the charge located in the center of the sphere. The diameter of the particle is given by $a\zeta$, where $a$ is a nearest neighbor lattice distance and $\zeta$ is the discretization parameter: see Fig. 1. Thus, in the special case $\zeta=1$ each ion  occupies a single site, while $\zeta=\infty$ corresponds to a continuum model. All particles interact only through  Coulombic potential and hard-core exclusion. Then the free energy of the system can be written as a sum of two terms, $F=F^{id}+F^{el}$. It allows to calculate the thermodynamic properties of the system, such as  chemical potentials and the pressure, as given by\cite{FL96,KKF02}
\begin{equation}\label{eq.thermo}
\overline{f} \equiv - \frac{F}{k_{B}TV}, \quad \overline{\mu}_{i} \equiv \frac{\mu}{k_{B}T}=-\frac{\partial \overline{f}}{\partial \rho_{i}}, \quad \overline{P} \equiv \frac {P}{k_{B}T}= \overline{f} +\sum_{i} \rho_{i} \overline{\mu}_{i}.
\end{equation}  

Since every positively charged  particle has a corresponding negatively charged ion, we can introduce a new hypothetical particles with chemical potential $\overline{\mu}_{1}=\overline{\mu}_{-}+\overline{\mu}_{+}$. Then the two-component system is easily mapped into the one-component system with overall density of hypothetical particles given by $\rho_{1}$. The conditions for phase transitions are specified by
\begin{equation}
\overline{\mu}_{1} (T,\rho_{liq}) = \overline{\mu}_{1} (T,\rho_{vap}), \quad \overline{P}(T,\rho_{liq})= \overline{P}(T,\rho_{vap}).
\end{equation}  

Let us take an arbitrary  particle and assume that its charge $q_{i}$ is  in  the center of coordinates. Then all lattice sites in the system can be divided into three groups with respect to the position of this fixed particle  as shown in Fig. 1. The first one consists of just one lattice cite, $\mathbf {r}=(0,0,0)$, where the charge of the particle is located. The second group is made of other lattice sites in the exclusion zone where no centers of other particles can be found. All lattice sites outside the exclusion zone are in the third group. To determine the average electrostatic potential at each lattice site we can derive the linearized lattice Poisson-Boltzmann equations following the standard  Debye-H\"{u}ckel procedures.\cite{mcquarrie,KKF02} For the central lattice site we obtain
\begin{equation} \label{eq1}
\Delta \varphi ( {\mathbf r}=(0,0,0) )= - \frac{4 \pi}{D v_{0}} q_{i},
\end{equation}  
with $v_{0}$ being a unit cell volume. In this equation $\Delta \varphi$ is the lattice Laplacian that defined as follows,\cite{KKF02}
\begin{equation}
\Delta \varphi ({\mathbf r})= - \frac{6 }{c_{0}a^{2}} \sum_{nn} [ \varphi({\mathbf r}+{\mathbf a})-\varphi({\mathbf r})],
\end{equation}  
where ${\mathbf a}$ is a nearest-neighbor vector and the summation runs over all $c_{0}$ nearest neighbors.

The number of the lattice sites in the exclusion zone (the second group) is determined by the discretization parameter $\zeta$ and the type of the lattice. For example, for sc lattices with $\zeta=1$, 2, 3, 4 or 5 these numbers are 0, 26, 92, 250 and 484, correspondingly.\cite{panagiotopoulos99} Since there are no charges in the exclusion zone outside of the central lattice site, the linearized  lattice Poisson-Boltzmann equations have a simple form,\cite{mcquarrie,KKF02}
\begin{equation}\label{eq2}
\Delta \varphi ({\mathbf r}_{ex})= 0,
\end{equation}
where ${\mathbf r}_{ex}$ specifies the lattice sites in the exclusion zone but not the central point.

Similarly, for the lattice sites in the third group, i.e., outside of the exclusion zone, the linearized  lattice Poisson-Boltzmann equations are given by\cite{mcquarrie,KKF02}
\begin{equation}\label{eq3}
\Delta \varphi ({\mathbf r}_{out})= \kappa^{2} \varphi ({\mathbf r}),
\end{equation}
where ${\mathbf r}_{out}$ describes the lattice points outside the exclusion zone, and $\kappa^{2}=4\pi\beta\rho_{1} q_{0}^{2}/D$ is the inverse squared Debye screening length  with $\beta=1/k_{B}T$.

Linearized  lattice Poisson-Boltzmann Eqs. (\ref{eq1}), (\ref{eq2}) and (\ref{eq3}) can be written in a unified way as
\begin{equation}\label{eq.general}
\Delta \varphi ({\mathbf r})= \kappa^{2} \varphi ({\mathbf r})-A_{0} \delta({\mathbf r})-\sum_{{\mathbf r}_{i,ex}} A_{i} \delta({\mathbf r}-{\mathbf r}_{i,ex}),
\end{equation}
where the coefficients $A_{0}$ and $A_{i}$ are determined explicitly from Eqs. (\ref{eq1}), (\ref{eq2}) and (\ref{eq3}), and the summation is over all lattice sites from the second group. Note that $\delta({\mathbf r})$ is a lattice delta-function defined as
\begin{equation}
\delta({\mathbf r}-{\mathbf r_{0}})= \left\{ \begin{array} {rc}
                                        1, &  \quad  {\mathbf r}={\mathbf r_{0}};    \\
                                        0, &  \quad   {\mathbf r} \ne {\mathbf r_{0}}.      
                                              \end{array}    \right.
\end{equation}
Eq. (\ref{eq.general}) can be solved via Fourier transformation and it yields the average electrostatic potential $\varphi ({\mathbf r})$ at every point of the lattice. It allows then to calculate the average electrostatic potential at the central site due to all ions except the one fixed at the origin,
\begin{equation}\label{psi}
\psi_{i}=\varphi_{i} ({\mathbf r},x^{2})-\varphi_{i} ({\mathbf r},x^{2}=0),
\end{equation}
where we introduced a dimensionless parameter $x=\kappa a$. It can be shown that the general expression for $\psi_{i}$ is given by
\begin{equation}
\psi_{i}= \frac{4\pi q_{i}}{D v_{0}} G(x^{2}),
\end{equation}
where the specific functional form of  $G(x^{2})$ is determined again from  Eqs. (\ref{eq1}), (\ref{eq2}) and (\ref{eq3}).

Since the ideal part of free energy of the lattice model of charged particles is known,\cite{KKF02} we only have to calculate the electrostatic contribution. It can be done by utilizing the Debye charging procedure,\cite{mcquarrie,KKF02}  
\begin{equation}\label{charging}
\overline{f}^{el}=-\frac{1}{k_{B}TV} \sum q_{j} \int_{0}^{1} \psi_{j}(\lambda q) d q= - \frac{1}{12 v_{0}} \int_{0}^{x^{2}} G(x^{2}) d (x^{2}).
\end{equation}
All thermodynamic properties and phase behavior of the system can be obtained from the free energy through standard calculations.\cite{FL96,KKF02}

\section{Lattice Electrolytes  with Different Discretization Parameters}

\subsection{Pure DH Theory for sc, bcc and fcc Lattices with $\zeta=1$}

Let us discuss in more detail the three-dimensional lattices with the smallest value of the discretization parameter, $\zeta=1$. This corresponds to the situation when each particle occupies one lattice site. Because in this case there are no lattice sites in the exclusion zone, the  linearized  lattice Poisson-Boltzmann equation (\ref{eq.general}) can be written in the following form,
\begin{equation}\label{eq.pureDH}
\Delta \varphi ({\mathbf r})= \kappa^{2} \varphi ({\mathbf r})-A_{0} \delta({\mathbf r}).
\end{equation}
The solution of this equation is found via Fourier transformation,
\begin{equation}\label{eq_A0}
\varphi ({\mathbf r})= A_{0} \frac{a^{2}}{6} \int_{k} \frac{ e^{i {\mathbf k}\cdot{\mathbf r}}}{(x^{2}+6)/6 - J({\mathbf k})},
\end{equation}
where $\int_{k} \equiv (2 \pi)^{-3} \int_{-\pi}^{\pi} d^{3} {\mathbf k}$, and  we defined a new lattice function $J({\mathbf k})$,
\begin{equation}
 J({\mathbf k})= \frac{1}{c_{0}} \sum_{nn} e^{i {\mathbf k} \cdot {\mathbf r}}.
\end{equation}
The coefficient $A_{0}$ can be found with the help of Eq. (\ref{eq1}),
\begin{equation} 
\Delta \varphi ({\mathbf r}=(0,0,0))= - \frac{4 \pi q_{i}}{D v_{0}} =\kappa^{2} \varphi ({\mathbf r}=(0,0,0))-A_{0}=A_{0} \frac{a^{2}}{6} \int_{k} \frac{1}{(x^{2}+6)/6 - J({\mathbf k})} - A_{0}.
\end{equation}  

Defining the integrated lattice Green's function as
\begin{equation}
P(z)=\int_{k} \frac{1}{1-z J({\mathbf k})},
\end{equation}
we obtain from Eq. (\ref{eq_A0})
\begin{equation}
A_{0}=\frac{4 \pi q_{i}}{D v_{0}} \ \frac{1}{1-\frac{x^{2}}{x^{2}+6} P\left( \frac{6}{x^{2}+6}\right)}.
\end{equation}
Thus  the average electrostatic potential at the origin is given by
\begin{equation}\label{eq.psi1}
\psi_{i}=\frac{4 \pi q_{i}}{D v_{0}} \frac{a^{2}}{6} \left[ \frac{1}{1-\frac{x^{2}}{x^{2}+6} P\left(\frac{6}{x^{2}+6}\right)} \ \frac{6}{x^{2}+6} P\left( \frac{6}{x^{2}+6}\right) - P(1) \right],
\end{equation}
from which, after applying the Debye charging procedure as outlined above, the electrostatic free energy density $\overline{f}^{el}$ can be easily calculated. The ideal lattice gas contribution to the free energy density is known,\cite{KKF02}
\begin{equation} \label{eq.ideal}
\overline{f}^{id}=-\frac{\rho_{1}^{*}}{v_{0}} \ln \rho_{1}^{*}-\frac{1-\rho_{1}^{*}}{v_{0}} \ln (1-\rho_{1}^{*}),
\end{equation}
where $\rho_{1}^{*}=\rho_{1} v_{0}$ is a reduced density of the free ions. 

Now we can perform explicit calculations for  the  ionic systems on sc, bcc and fcc three-dimensional lattices. The lattice functions $J({\mathbf k})$ are given by\cite{KKF02}
\begin{equation} 
J({\mathbf k})= \left\{ \begin{array} {ll}
                       \frac{1}{3} (\cos k_{1} +\cos k_{2} + \cos k_{3}) & \quad (sc),\\
                       \cos k_{1} \cos k_{2} \cos k_{3} & \quad (bcc),\\
                       \frac{1}{3}(\cos k_{1}\cos k_{2} +\cos k_{2}\cos k_{3} + \cos k_{1}\cos k_{3}) & \quad (fcc),
                       \end{array} \right.
\end{equation}
with $-\pi \le k_{1}, k_{2}, k_{3} \le \pi$. The corresponding integrated lattice Green's functions can be evaluated numerically exactly   by using the fact that they can be expressed in terms of elliptic integrals.\cite{joyce98,KKF02}

The thermodynamic properties and phase behavior of lattice electrolytes on sc, bcc and fcc lattices are  explicitly calculated from Eqs. (\ref{eq.thermo}), (\ref{eq.psi1}) and (\ref{eq.ideal}). The predicted phase coexistence curves are plotted in Fig. 2, while the critical parameters are given in Table 1. To analyze the effect of different discretizations we also present in Fig. 2 and Table 1 the phase diagram and critical parameters for the pure DH model of continuum ($\zeta=\infty$) hard-core ionic fluids.\cite{FL96} It can be seen that the critical temperatures for lattice models are approximately $30\%$ higher than the corresponding continuum value. At the same time  the thermodynamic properties and critical parameters  of the  lattice pure DH models are approaching the continuum model as the number of nearest neighbors is growing from $c_{0}=6$ (sc lattice ) to $c_{0}=12$ (fcc lattice), as expected. It also should be noted that in the lattice models the  predicted phase coexistence curves have a physically reasonable behavior in the limit of high density and $T \rightarrow 0$, in contrast to the original lattice DH theories.\cite{KKF02} This is due to the fact that the presented theoretical method correctly  describes the average potential $\psi$  at all densities of charged particles.

\subsection{Pure DH Theory for sc Lattice with $\zeta=\sqrt 2, \sqrt 3$ and 2}

Now we can examine the lattice models of electrolytes with discretization parameters $\zeta >1$. Thermodynamic properties will be expressed in terms of the reduced density and the reduced temperature, namely,
\begin{equation}
\rho_{1}^{*}=(a \zeta)^{3} \rho_{1}, \quad \mbox{and} \quad  T^{*}= \frac{k_{B}TD a \zeta}{q_{0}^{2}}.
\end{equation}
Consider first the ionic lattice model with $\zeta= \sqrt 2$. In this case the exclusion zone around any charged particle consists of the lattice central site, where the charge is located, and 6 nearest-neighbor lattice sites where no other charges can be found due to geometrical constraints --- see Fig. 1. Then the  linearized  lattice Poisson-Boltzmann equation (\ref{eq.general}) yields
\begin{equation}
\Delta \varphi ({\mathbf r})= \kappa^{2} \varphi ({\mathbf r})-A_{0} \delta({\mathbf r})- A_{1} \sum_{nn} \delta({\mathbf r}-{\mathbf a}_{nn}),
\end{equation}
where ${\mathbf a}_{nn}$ described the position of the lattice sites in the exclusion zone outside of the central site. The solution of this equation can be found again using  the Fourier transformation, and it is given by
\begin{equation}\label{eq.phi}
\varphi ({\mathbf r})=A_{0} \frac{a^{2}}{6} \int_{k} \frac{ e^{i {\mathbf k}\cdot{\mathbf r}}}{(x^{2}+6)/6 - J({\mathbf k})} + 6 A_{1} \frac{a^{2}}{6} \int_{k} \frac{ J({\mathbf k}) e^{i {\mathbf k}\cdot{\mathbf r}}}{(x^{2}+6)/6 - J({\mathbf k})}.
\end{equation}
The boundary conditions to determine the coefficients $A_{0}$ and $A_{1}$ are found from Eqs. (\ref{eq1}) and (\ref{eq2}):
\begin{equation}\label{eq.conditions}
-\frac{4 \pi q_{i}}{Dv_{0}} =\kappa^{2} \varphi ({\mathbf r}=(0,0,0))- A_{0}, \quad 0=\kappa^{2} \varphi ({\mathbf r}=(1,0,0))-A_{1}.
\end{equation}
From Eqs. (\ref{eq.phi}) and (\ref{eq.conditions}) we derive
\begin{equation}
A_{0}=\frac{4 \pi q_{i}}{D v_{0}} \ \frac{\left[ (6+x^{2})(-1+\frac{1}{6}x^{2}(6+x^{2})(-1+P\left( \frac{6}{x^{2}+6}\right) \right]}{-(1+x^{2})(6+x^{2})+x^{2}(7+x^{2}) P\left( \frac{6}{x^{2}+6}\right)},
\end{equation}
\begin{equation}
A_{1}=-\frac{4 \pi q_{i}}{D v_{0}} \ \frac{\left[ \frac{1}{6}x^{2}(6+x^{2})(-1+P\left( \frac{6}{x^{2}+6}\right) \right]}{-(1+x^{2})(6+x^{2})+x^{2}(7+x^{2}) P\left( \frac{6}{x^{2}+6}\right)}.
\end{equation}
Thus the average electrostatic potential at the origin due to all other ions, defined in Eq. (\ref{psi}), again can be expressed in terms of the integrated lattice Green's function $P(z)$,
\begin{equation}
\psi_{i}=\frac{4 \pi q_{i}}{D v_{0}} \frac{a^{2}}{6} \left[ A_{0}' \left( \frac{x^{2}}{x^{2}+6} P \left( \frac{6}{x^{2}+6} \right) - P(1) \right) + 6 A_{1}' \left( P \left(\frac{6}{x^{2}+6} \right) - P(1) \right) \right],
\end{equation}
where $A_{j}'= \frac{Dv_{0}}{4 \pi q_{i}} A_{j}$ for $j=0$ and 1. After Debye charging procedure, as described above in Eq. (\ref{charging}), this equation allows us to calculate the electrostatic contribution to thermodynamic properties. 

The exact expressions for the ideal free energy contributions for the lattice models with $\zeta >1$ are not known, but for low-densities the expression (\ref{eq.ideal}) is still approximately valid. Then the thermodynamic properties and phase behavior of the system can be calculated in a similar way  as was done for $\zeta=1$ lattices. The critical parameters for the sc lattice model of electrolytes with $\zeta=\sqrt 2$ are $T_{c}^{*}=0.069$ and $ \rho_{c}^{*}=0.0054$. 

For the system of charged particles on the lattice with the discretization parameter $\zeta = \sqrt 3$ the exclusion zone around any arbitrary ion consists of 19 lattice sites: 1 central site, 6 nearest-neighbor sites, and 12 next-nearest-neighbor sites. For this model the general  linearized  lattice Poisson-Boltzmann equation (\ref{eq.general}) has the following form
\begin{equation}
\Delta \varphi ({\mathbf r})= \kappa^{2} \varphi ({\mathbf r})-A_{0} \delta({\mathbf r})- A_{1} \sum_{nn} \delta({\mathbf r}-{\mathbf a}_{nn})-A_{2} \sum_{nnn} \delta({\mathbf r}-{\mathbf a}_{nnn}),
\end{equation}
with ${\mathbf a}_{nnn}$ defining the positions of next-nearest-neighbor lattice sites. As before, we can determine the coefficients $A_{0}$, $A_{1}$ and $A_{2}$ with the help of Eqs. (\ref{eq1}) and (\ref{eq2}),
\begin{equation}
-\frac{4 \pi q_{i}}{Dv_{0}} =\kappa^{2} \varphi ({\mathbf r}=(0,0,0))- A_{0}, \quad 0=\kappa^{2} \varphi ({\mathbf r}=(1,0,0))-A_{1}, \quad 0=\kappa^{2} \varphi ({\mathbf r}=(1,1,0))-A_{2}.
\end{equation} 
Unfortunately, for the lattice models with $\zeta > \sqrt 2$ the electrostatic part of the reduced free energy cannot be expressed only in terms of the integrated lattice Green's function. However, for explicit numeric evaluations of the thermodynamic properties the following equality can be used\cite{katsura71,kobelev03} 
\begin{equation}
\int_{k} \frac{e^{i{\mathbf k} \cdot (lx,my,nz)}}{\alpha - J({\mathbf k})} = 3 \int_{0}^{\infty} dt e^{-3 \alpha t} I_{l}(t) I_{m}(t) I_{n}(t),
\end{equation} 
where $I_{k}(t)$ is a modified Bessel function of the first kind. Then the thermodynamic calculations indicate that the critical parameters for the  system of charged particles on the lattice with $\zeta= \sqrt 3$ are given by $T_{c}^{*}=0.063$ and $ \rho_{c}^{*}=0.0047$.

Similar procedure  can be performed for the lattice model of electrolytes  with the discretization parameter $\zeta=2$. The schematic picture for this model is shown in Fig. 1. In this case there are 27 lattice sites in the exclusion zone. The general  linearized  lattice Poisson-Boltzmann equation (\ref{eq.general}) can be written as 
\begin{equation}
\Delta \varphi ({\mathbf r})= \kappa^{2} \varphi ({\mathbf r})-A_{0} \delta({\mathbf r})- A_{1} \sum_{nn} \delta({\mathbf r}-{\mathbf a}_{nn})-A_{2} \sum_{nnn} \delta({\mathbf r}-{\mathbf a}_{nnn})-A_{3} \sum_{nnn} \delta({\mathbf r}-{\mathbf a}_{nnn}).
\end{equation}
The solution of this equation leads to determination of average electrostatic potential that allows to obtain the electrostatic contribution to  free energy density. The resulting thermodynamic calculations yield the following values of critical parameters for this model: $T_{c}^{*}=0.066$ and $ \rho_{c}^{*}=0.0060$.

Critical parameters of lattice models of ionic systems with different discretizations, obtained in pure DH theory, are shown in Fig. 3. It can be seen that the increase in the discretization parameter $\zeta$ generally lowers the values of critical temperature and density, although the dependence is non-monotonic. This noisy behavior of critical parameters, especially for non-integer $\zeta$, has also  been observed in Monte Carlo computer simulations of lattice models with large $\zeta \ge 5$.\cite{kim04}

\subsection{Bjerrum Ion Pairing for Lattices with $\zeta=1$}

It is known that the pure Debye-H\"{u}ckel  theory, that takes into account only free ions, is not successful for the description of thermodynamic properties of electrolytes at low temperatures.\cite{KKF02,FL96} At these conditions the positive and negative particles have a tendency to stick together in ion pairs. The framework of the lattice description of ionic systems is very convenient for  complete  analysis of the process of ion pairing, and it avoids the problem encountered in the continuum models of electrolytes.\cite{FL96,KKF02}

Here we consider the pairing as a reversible chemical reaction of the association of positive and negative ions and the production of neutral pairs. This process is controlled by equilibrium constant $K(T)$ that can be determined from the densities of different species. Ion pairs are specified by a density $\rho_{2}$ and a chemical potential $\mu_{2}$. In the simplest Debye-H\"{u}ckel-Bjerrum (DHBj) approximation  we neglect the Coulombic interactions between ion pairs and free charged particles.\cite{KKF02,FL96} 

The condition of chemical equilibrium between the ion pairs and free ions means that $\mu_{2}=\mu_{+}+\mu_{-}=\mu_{1}$. Let us introduce thermodynamic activities of the particles in the system:
\begin{equation}\label{activities}
z_{1}=2z_{+}=2z_{-}, \quad z_{1}=\frac{2\theta_{1}}{\Lambda_{1}^{3}} e^{\overline{\mu}_{1}}, \quad z_{2}=\frac{\theta_{2}}{\Lambda_{2}^{6}} e^{\overline{\mu}_{2}},
\end{equation}
where $\Lambda_{i}$ denotes the de Broglie wavelengths, and $\Lambda_{+}=\Lambda_{-}=\Lambda_{1}=\Lambda_{2}$; the parameters $\theta_{i}$ define the appropriate internal configurational partition functions.\cite{FL96,KKF02} The chemical equilibrium condition can be expressed in the form of the mass-action law, $z_{2}=\frac{1}{4} K z_{1}^{2}$, which yields for equilibrium constant $K(T)$,
\begin{equation}
K(T)=\theta_{2}(T)=\sum_{nn}e^{-\beta q_{0} \varphi ({\mathbf a}_{nn})}=c_{0} e^{-\beta q_{0} \varphi ({\mathbf a}_{nn})}.
\end{equation}
With the help of  Widom's potential-distribution theorem,\cite{widom63,KKF02} the chemical potential for free ions is given by
\begin{equation}\label{eq.mu1}
\overline{\mu}_{1}=\ln \left( \frac{\rho_{1}^{*}}{1-\rho_{1}^{*}-2 \rho_{2}^{*}} \right) + \ln \left( \frac{\Lambda_{1}^{3}}{\theta_{1}} \right) + \overline{\mu}_{1}^{el}.
\end{equation}
To obtain the corresponding expression for the chemical potential of ion pairs we use the Bethe approximation. \cite{KKF02,artyomov03,nagle66} In this case it yields
\begin{equation}
z_{2}=\frac{(2 \rho_{2}^{*}/c_{0}) \left[ 1- (2 \rho_{2}^{*}/c_{0})  \right]}{(1-\rho_{1}^{*}-2 \rho_{2}^{*})^{2}},
\end{equation}
and, finally,
\begin{equation}\label{eq.mu2}
\overline{\mu}_{2}=\ln \left( \frac{(2 \rho_{2}^{*}/c_{0}) \left[ 1- (2 \rho_{2}^{*}/c_{0})  \right]}{(1-\rho_{1}^{*}-2 \rho_{2}^{*})^{2}} \right) + \ln \left( \frac{\Lambda_{2}^{6}}{\theta_{2}} \right).
\end{equation}
Using the expression for the mass-action law and Eqs. (\ref{activities}), (\ref{eq.mu1}) and (\ref{eq.mu2}), we obtain
\begin{equation}
\rho_{2}^{*}=\frac{c_{0}}{4} \left[ 1- \sqrt{1-c_{0}(\rho_{1}^{*})^{2}  \exp \left\{ \frac{2 \pi a^{3}}{3 T^{*} v_{0}} \left[ \frac{P(\frac{6}{x^{2}+6})-1}{1-\frac{x^{2}}{x^{2}+6} P(\frac{6}{x^{2}+6})} \right] \right\} } \right].
\end{equation}
This expression specifies the density of  ion pairs in terms of densities of free charged particles. 

Because the ion pairs are neutral, in  DHBj approximation they do not contribute into the electrostatic free energy. Thus, 
\begin{equation}\label{eq.ideal1}
\overline{f}=\overline{f}^{id}(\rho_{1}^{*},\rho_{2}^{*}) + \overline{f}^{el} (\rho_{1}^{*}),
\end{equation}
where the ideal contribution to free energy is given by\cite{KKF02}
\begin{equation}
\overline{f}^{id}(\rho_{1}^{*},\rho_{2}^{*}) = \rho_{1}^{*} \ln \rho_{1}^{*} - (1-\rho_{1}^{*}-2\rho_{2}^{*}) \ln (1-\rho_{1}^{*}-2\rho_{2}^{*}) - \rho_{2}^{*} \ln \rho_{2}^{*} - (\rho_{2}^{*}-3) \ln (1-2\rho_{2}^{*}/c_{0}).
\end{equation}
The phase coexistence curve for sc lattice in DHBj approximation is shown in Fig. 4. The critical temperature is slightly lower, while the critical density is much larger than in the pure DH approximation. But it also has an unphysical banana-like shape, and this is because of the neglect of electrostatic interactions of ion pairs with free charged particles, as explained for continuum and lattice models of electrolytes.\cite{FL96,KKF02}

\subsection{Dipole-Ion Interactions for Lattices with $\zeta=1$}

Clearly unphysical phase behavior in DHBj approximation is due to the fact that the properties of ion pairs as dipoles  are not taken into account. As was shown earlier,\cite{FL96,KKF02} for the realistic description of ionic fluids it is important to consider the ion pairs as dipoles that interact with residual free charged particles. These solvation effects eliminate the unphysical phase behavior and provide a better agreement between the calculated critical parameters and the parameters estimated from the computer simulations.\cite{FL96,KKF02}     

Let us consider an arbitrary dipole particle that occupies 2 neighboring lattice sites. Let assume that the positive charge of the particle is at the site $(0,0,0)$, while the negative charge is at $(1,0,0)$. The  linearized  lattice Poisson-Boltzmann equation (\ref{eq.general}) in this case can be written in the following form,
\begin{equation}\label{eq.DHBjDI}
\Delta \varphi ({\mathbf r})= \kappa^{2} \varphi ({\mathbf r})-A_{0} \delta({\mathbf r})- A_{1} \delta({\mathbf r}-(1,0,0)).
\end{equation}
Comparing  with Eq. (\ref{eq.pureDH}), we note that the last 2 terms in this expression reflect the fact that the particle is a dipole, consisting of two opposite charges. The Eq. (\ref{eq.DHBjDI}) can be solved as before,
\begin{equation}
\varphi ({\mathbf r})=  \frac{a^{2}}{6} \int_{k} \frac{A_{0}+A_{1} e^{i {\mathbf k}\cdot(-1,0,0)}} {(x^{2}+6)/6 - J({\mathbf k})} e^{i {\mathbf k} \cdot{\mathbf r}},
\end{equation}
with coefficients $A_{0}$ and $A_{1}$ determined from the following conditions:
\begin{equation}
\Delta \varphi ({\mathbf r}=(0,0,0)) = -\frac{4 \pi q_{+}}{ D v_{0}} = \kappa^{2} \varphi ({\mathbf r}=(0,0,0))-A_{0},
\end{equation}
and
\begin{equation}
\Delta \varphi ({\mathbf r}=(1,0,0)) = -\frac{4 \pi q_{-}}{ D v_{0}} = \kappa^{2} \varphi ({\mathbf r}=(1,0,0))-A_{1}.
\end{equation}
These equations lead to the expression for the average electrostatic potential $\psi_{i}^{DI}$ due to all ions except the positive ion fixed at $(0,0,0)$ and negative ion fixed at  $(1,0,0)$, 
\begin{equation}
\psi_{i}^{DI}= \frac{4 \pi q_{i}}{ D v_{0}} \ \frac{a^{2}}{6} \left[ \frac{x^{2}(1- P\left( \frac{6}{x^{2}+6}\right))}{6-x^{2}+\frac{x^{4}}{x^{2}+6} P\left( \frac{6}{x^{2}+6}\right)} \right] = \frac{4 \pi q_{i}}{ D v_{0}} \ \frac{a^{2}}{6} \ G(x^{2}),   
\end{equation}
where we defined an auxiliary function $G(x^{2})$,
\begin{equation}
G(x^{2}) =  \frac{x^{2}(1- P\left( \frac{6}{x^{2}+6}\right))}{6-x^{2}+\frac{x^{4}}{x^{2}+6} P\left( \frac{6}{x^{2}+6}\right)}. 
\end{equation}

The contribution to the free energy density due to the interactions between dipoles and free ions can be calculated from
\begin{equation}
\overline{f}^{DI} = - \frac{ 2 \pi \rho_{2}^{*}}{3 T^{*} v_{0}} \ \frac{1}{x^{2}} \ \int_{0}^{x^{2}} G(x^{2}) d x^{2}. 
\end{equation}
The overall free energy density of the system can be written as
\begin{equation}
\overline{f}=\overline{f}^{id}(\rho_{1}^{*},\rho_{2}^{*}) + \overline{f}^{el} (\rho_{1}^{*}) +\overline{f}^{DI},
\end{equation}
where $ \overline{f}^{id}(\rho_{1}^{*},\rho_{2}^{*}) $ is given by Eq. (\ref{eq.ideal1}).  From these equations all thermodynamic properties can be obtained.

The resulting phase coexistence curves and the critical parameters for sc, bcc and fcc lattices estimated using the full  DHBjDI approach are presented  in Fig. 4  and Table 2, respectively. The predicted critical temperatures  are decreasing as the number of nearest neighbors is going up, but they are still 25-30 $\%$ higher than the corresponding values for the continuum DHBjDI theory\cite{FL96} and for the Monte carlo simulations.\cite{caillol02,panagiotopoulos02,kim04} At the same time, the critical densities are approaching the predictions of continuum RPM calculations, while deviating significantly from the computer simulations estimates.\cite{FL96,caillol02,panagiotopoulos02}

\section{Summary and Conclusions}

A theoretical investigation of ionic  systems  on the lattices with different discretizations is presented. A general lattice theory,  based on the  Debye-H\"{u}ckel approach, is developed for lattices with  any discretization parameter. The theory is applied for explicit calculations of thermodynamic properties, phase behavior and critical parameters of several lattice models of electrolytes.  

The simplest version of general lattice theory, the  pure  Debye-H\"{u}ckel theory that takes into account only the free ions, is utilized for obtaining the thermodynamic properties of sc, bcc and fcc lattices for the discretization parameter $\zeta=1$, and for sc lattices with $\zeta= \sqrt 2$, $\sqrt 3$ and $2$. All  considered cases exhibit the low-density gas-liquid phase transitions. As expected, with the increase in the lattice discretization parameter  $\zeta$ the critical parameters are generally decreasing and approaching the values obtained in the continuum RPM treatment of ionic fluids\cite{FL96} that corresponds to $\zeta=\infty$ case. The behavior of the critical parameters is non-monotonic, especially for non-integer $\zeta$, in agreement with observations from latest Monte Carlo computer simulations  for large discretization parameters ($\zeta \ge 5$).\cite{kim04}

For ionic fluids on the sc, bcc and fcc lattices with $\zeta=1$ a more realistic theoretical description, that takes into account the creation of ion pairs and the ion-dipole interactions, is developed. The consideration of ion pairs as neutral species (the DHBj approach) slightly lowers the critical temperature and significantly increases the critical density, while the overall phase coexistence curves have unphysical banana-like shapes. This unphysical behavior is cured when the ion pairs are treated as dipoles that interact with the residual free charged particles. The resulting critical temperatures decrease even more, while the values of critical densities  are also lower. The critical temperatures for the lattice electrolytes are slightly higher than the continuum RPM values and the computer simulations estimates. The predicted critical densities for the lattice models are essentially the same as in the corresponding continuum models of electrolytes. However, they are much lower than the critical densities from Monte Carlo simulations. This behavior is typical for lattice models.   

Our theoretical approach allows to calculate the thermodynamic properties of ionic systems at low densities. However, at higher densities the sublattice ordering becomes very important in sc and bcc lattice models of electrolytes.\cite{dickman99,KKF02} The original lattice Debye-H\"{u}ckel approach\cite{KKF02} was able to capture the ordering processes. It was predicted that the sublattice ordering would suppress the gas-liquid phase coexistence and the tricritical point would appear, in accord with the recent lattice computer simulations.\cite{kobelev03}  The predicted tricritical density  was in agreement with the estimates from Monte Carlo computer simulations,\cite{kobelev03} while the critical temperature was overestimated. We plan to investigate the sublattice ordering using the modified lattice DH method, developed in this paper, in a near future.  In addition, our theoretical method might be used to investigate many other problems related to ionic fluids, such as the effect of additional short-range interactions, the size asymmetry and charge asymmetry, and the multi-component mixtures.

\section*{Acknowledgments}

The authors would like to acknowledge the support from the Welch Foundation (grant  C-1559), the Alfred P. Sloan foundation (grant  BR-4418)  and the U.S. National Science Foundation (grant CHE-0237105). The authors also are grateful to M.E. Fisher for valuable discussions and encouragements.

\newpage
\begin{table}[h]
\centering
\caption{ Critical parameters for lattice electrolytes  in the pure DH theory. HC corresponds to the predictions from the continuum RPM with hard-core interactions.\cite{FL96}}
\vspace{5mm}
\begin{tabular}{|c|c|c|} \hline
Model                 &   $T_{c}^{*}$                                     & $\rho_{c}^{*}$                                                    \\ \hline
sc                    & \hspace{5mm}  0.084377  \hspace{5mm}              & \hspace{5mm} 0.007978      \hspace{5mm}    \\
bcc                   & \hspace{5mm}  0.083793  \hspace{5mm}              & \hspace{5mm} 0.005659      \hspace{5mm}     \\ 
fcc                   & \hspace{5mm}  0.082415  \hspace{5mm}              & \hspace{5mm} 0.004591      \hspace{5mm}     \\ 
HC                    & \hspace{5mm}  0.061912   \hspace{5mm}             & \hspace{5mm} 0.004582       \hspace{5mm}    \\ \hline
\end{tabular}
\end{table}

\newpage
\begin{table}[h]
\centering
\caption{ Critical parameters from the full  DHBjDI theory for lattice and continuum electrolytes. For comparison, the estimates of Monte Carlo computer simulations are also shown.} 
\vspace{5mm}
\begin{tabular}{|c|c|c|} \hline
Model                                                 &   $T_{c}^{*}$                                     & $\rho_{c}^{*}$                         \\ \hline
sc                                                    & \hspace{5mm}  0.0756  \hspace{5mm}                & \hspace{5mm} 0.0422      \hspace{5mm}    \\
bcc                                                   & \hspace{5mm}  0.0752  \hspace{5mm}                & \hspace{5mm} 0.0336      \hspace{5mm}     \\ 
fcc                                                   & \hspace{5mm}  0.0686  \hspace{5mm}                & \hspace{5mm} 0.0292      \hspace{5mm}     \\ 
continuum\cite{FL96}                                  & \hspace{5mm}  0.0522-0.0554  \hspace{5mm}         & \hspace{5mm} 0.0244-0.0259   \hspace{5mm}     \\ 
simulations\cite{caillol02,panagiotopoulos02,kim04}   & \hspace{5mm}  0.049   \hspace{5mm}                & \hspace{5mm} 0.06-0.08      \hspace{5mm}    \\ \hline
\end{tabular}
\end{table}

\newpage

\noindent {\bf Figure Captions:} \\\\

\noindent Fig. 1 \quad  A schematic picture of the lattice model of electrolytes with different discretizations. Large empty circles denote the positive particles, while large gray circles correspond to the negative ions. The ions are spheres with the diameter $a \zeta$, where $a$ is the distance between the nearest lattice sites. The position of the origin is marked by a cross. Small filled circles indicate the position of lattice sites in the exclusion zone of the central positive ion.   

\vspace{5mm}

\noindent Fig. 2 \quad  Phase coexistence curves predicted by pure DH theory for lattice electrolytes with $\zeta=1$: (a) sc, (b) bcc, (c) fcc. The  predictions from the continuum model of electrolytes with hard-core interactions\cite{FL96} are shown in (d). 

\vspace{5mm}

\noindent Fig. 3 \quad Critical parameters  of lattice models of electrolytes for pure DH theory  as a function of inverse lattice discretization parameters $\zeta$: a) critical temperatures, b) critical densities.  

\vspace{5mm}

\noindent Fig. 4 \quad Phase coexistence curves of ionic systems  calculated for (a) sc lattice in the DHBj approximation; and using the full DHBjDI approach for (b) sc, (b) bcc, and (d) fcc lattices.

\newpage

\begin{figure}[ht]
\begin{center}
\vskip 1.5in
\unitlength 1in
\begin{picture}(4.0,4.0)
  \resizebox{3.375in}{3.2in}{\includegraphics{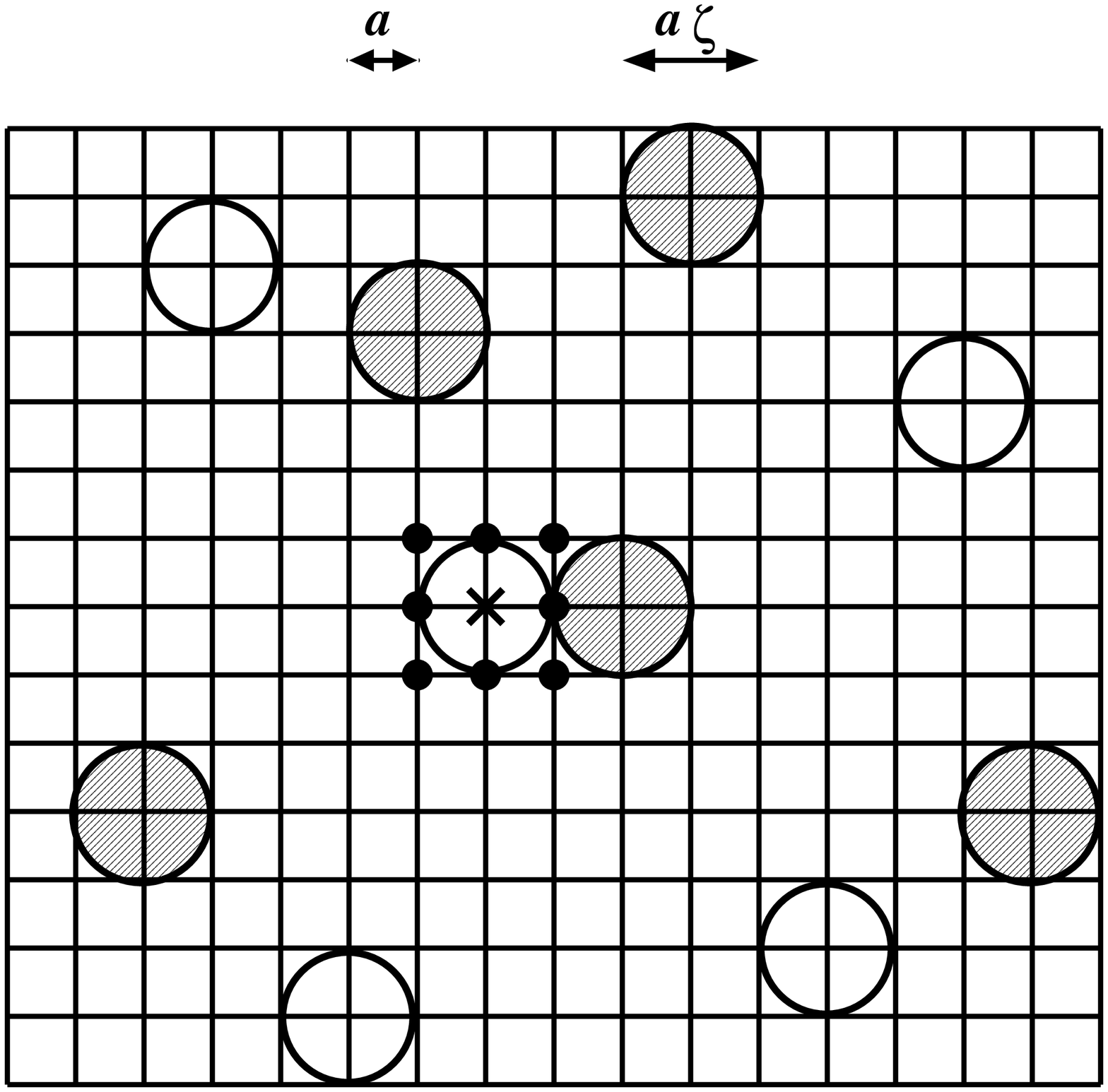}}
\end{picture}
\vskip 3in
 \begin{Large}  \end{Large}
\end{center}
\vskip 3in
\end{figure}

\newpage

\begin{figure}[ht]
\begin{center}
\vskip 1.5in
\unitlength 1in
\begin{picture}(4.0,4.0)
  \resizebox{3.375in}{3.1in}{\includegraphics{Fig2.eps}}
\end{picture}
\vskip 3in
 \begin{Large}  \end{Large}
\end{center}
\vskip 3in
\end{figure}

\newpage

\begin{figure}[ht]
\begin{center}
\vskip 1.5in
\unitlength 1in
\begin{picture}(4.0,4.0)
  \resizebox{3.375in}{3.1in}{\includegraphics{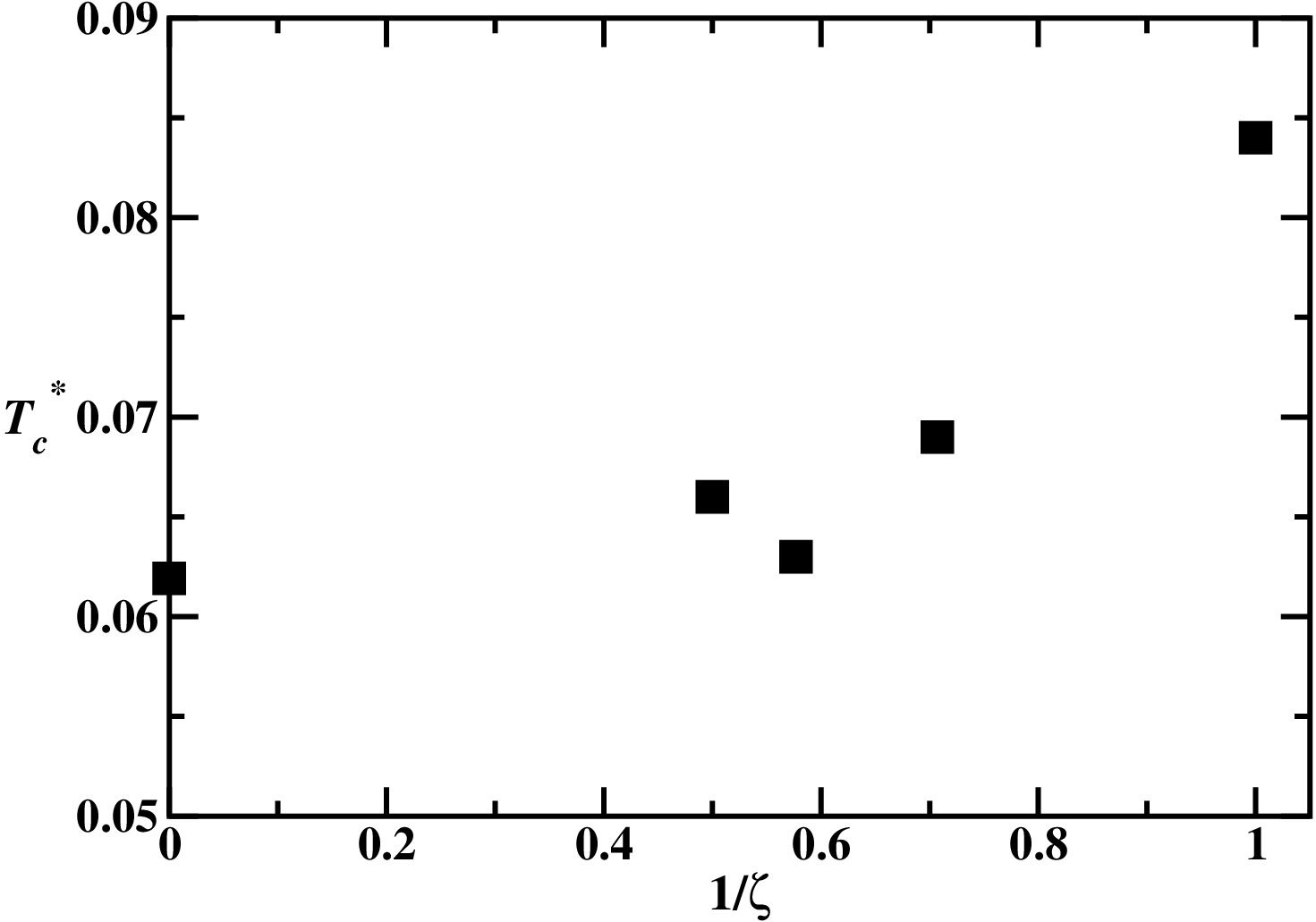}}
\end{picture}
\vskip 3in
 \begin{Large}  \end{Large}
\end{center}
\vskip 3in
\end{figure}

\newpage

\begin{figure}[ht]
\begin{center}
\vskip 1.5in
\unitlength 1in
\begin{picture}(4.0,4.0)
  \resizebox{3.375in}{3.1in}{\includegraphics{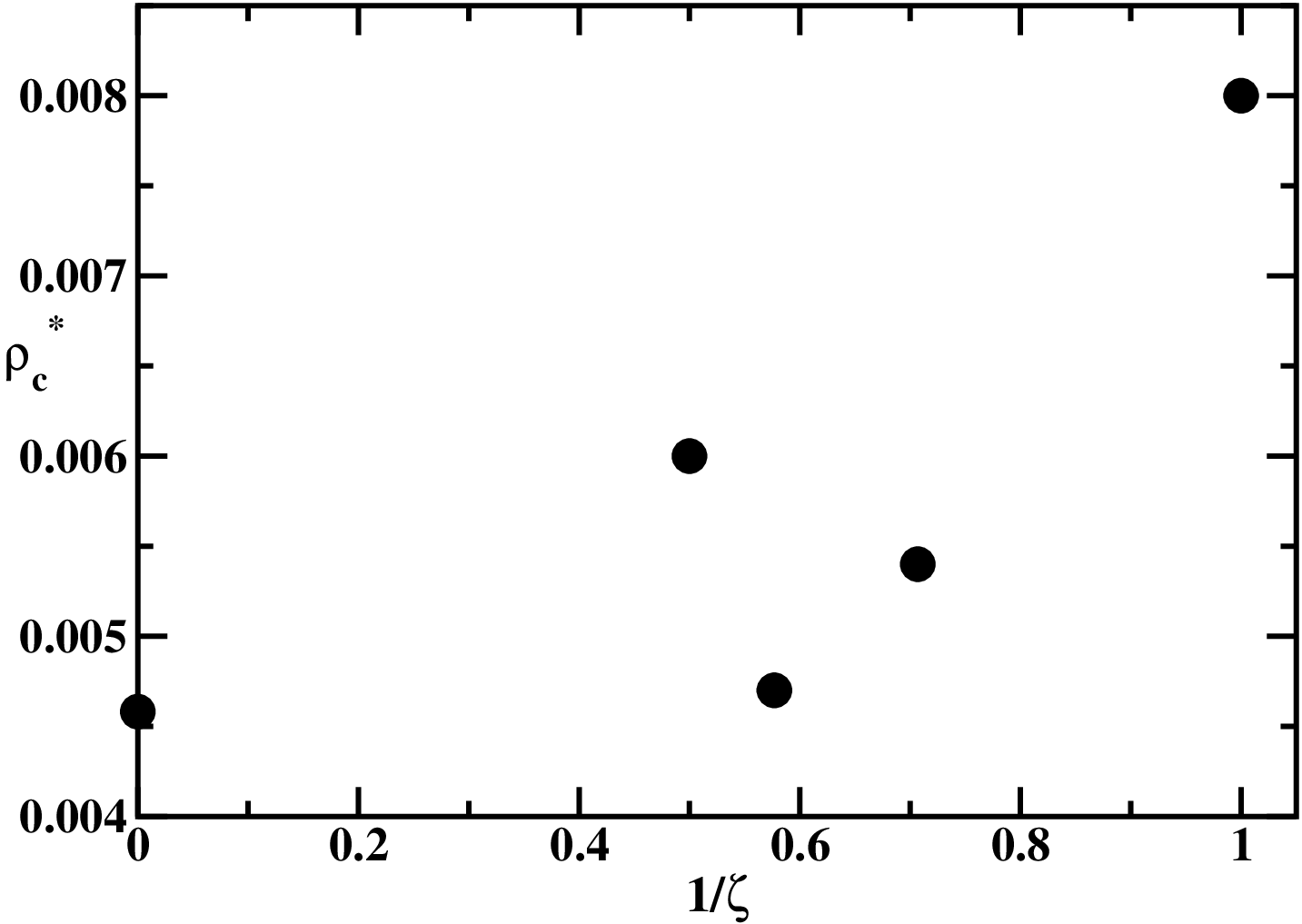}}
\end{picture}
\vskip 3in
 \begin{Large}  \end{Large}
\end{center}
\vskip 3in
\end{figure}

\newpage

\begin{figure}[ht]
\begin{center}
\vskip 1.5in
\unitlength 1in
\begin{picture}(4.0,4.0)
  \resizebox{3.375in}{3.1in}{\includegraphics{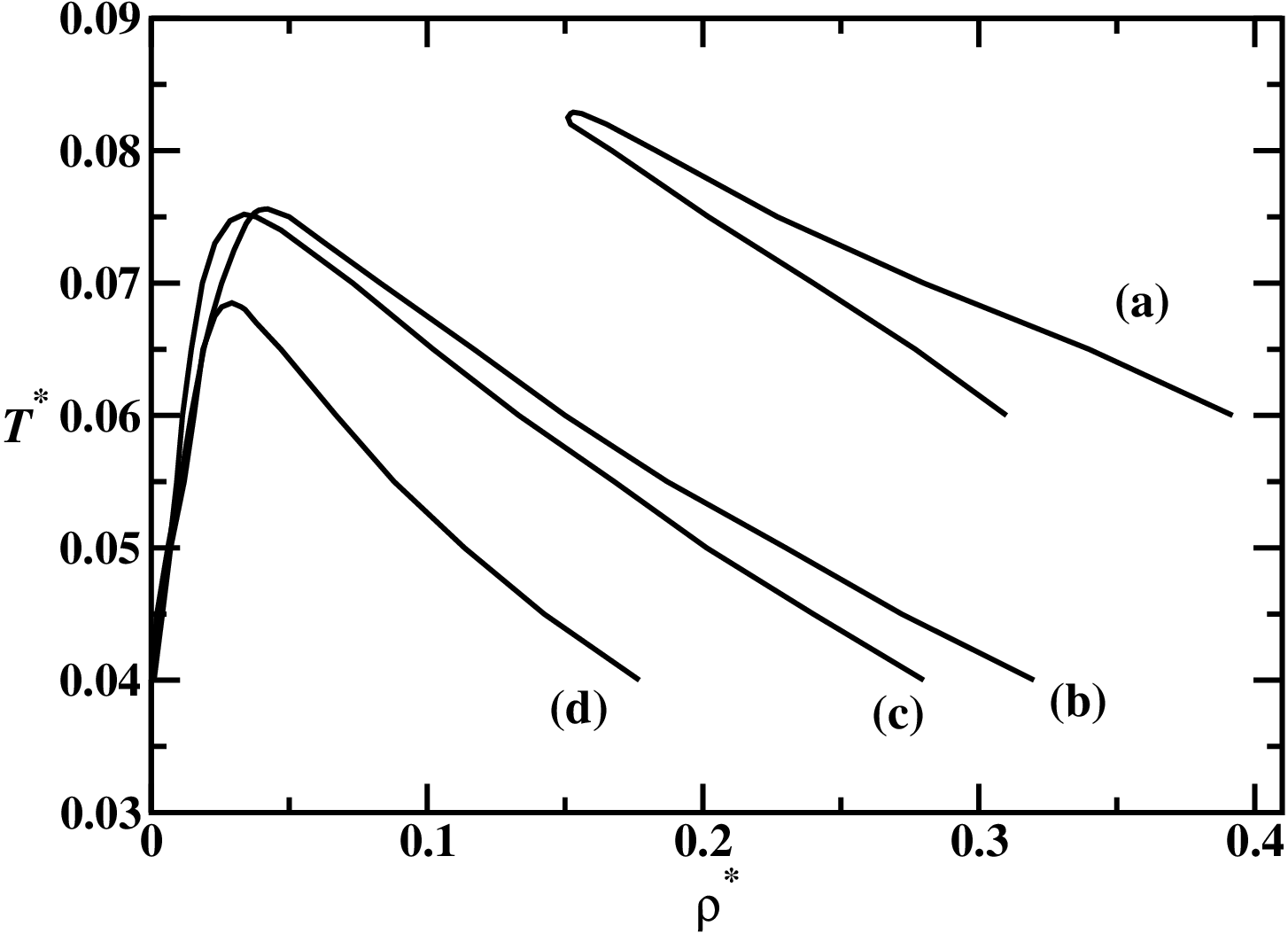}}
\end{picture}
\vskip 3in
 \begin{Large}  \end{Large}
\end{center}
\vskip 3in
\end{figure}

\end{document}